\documentclass[10pt,twoside]{book}
\usepackage{multirow}
\usepackage{graphicx}
\usepackage{xspace}  
\usepackage{lysa}
\usepackage{hero}   
\usepackage{graphicx}
\usepackage{makeidx}  
\usepackage{verbatim}

\def\thesisyear{2010} 
\def\thesisnumber{70}  
\def\thesisauthor{Ender Y\"uksel} 
\def\thesistitle{Analysing ZigBee Key Establishment Protocols} 
\def\thesiskeywords{mathematical modelling}
\def\thesisversion{print} 

\input{immphddef.sty}

\chapter{Preface}

This report provides a detailed documentation on the application of
static program analysis to the key establishment protocols of the
ZigBee wireless sensor network standard. The approach presented in this
report is within the scope of the SENSORIA (\emph{Software Engineering for
Service-Oriented Overlay Computers}) project, and will form a
preliminary version of one of the chapters in my PhD dissertation. The
discovered flaw and the proposed secure protocols were recently
published in a conference paper (see references \cite{yuk:nie:2009})
and also accepted for journal publication.

\paragraph{Acknowledgement} 
This work has been partially supported by EU-FETPI Global Computing
Project IST-2005-16004 SENSORIA.

I would like to thank Hanne Riis Nielson and Flemming Nielson for the
supervision and invaluable contribution in this work, and also for providing
me with this opportunity to work on an international research
project.

I would also like to thank Gavin Lowe for kind discussions and
feedback on fixing the flaw in the key establishment
protocol.

 \vspace{5mm} \mbox{}\hfill
\begin{minipage}[t]{180mm}
  Kongens Lyngby, February 2010
  \\ \\
  Ender Y\"uksel
\end{minipage}

\markboth{}{}

\setcounter{tocdepth}{2}
\tableofcontents
\clearpage
\listoftables

\mainmatter

\chapter{Analyzing the Protocols}
\label{chapter:3}
Computer networks or simply networks are the main means of information
sharing and communication in today's IT infrastructure. Certain
protocols are executed to facilitate communication in
networks. However, such networks are mostly insecure and the
communication needs to be protected against attackers that may
influence network traffic and and communication parties that might be either dishonest or compromised by attackers.

Cryptographic security protocols form an essential ingredient of
network communications by ensuring secure communication over insecure
networks. These protocols use cryptographic primitives to support
certain security properties, but ensuring this properties requires a
lot more effort. Despite the relatively small size of the security
protocols it is very hard to design them correctly, and their analysis
is very complicated. One of the most well-known examples is the
Needham-Schroeder protocol \cite{nee:sch}, that was proven secure by using  BAN
logic \cite{ban}. Seventeen years later G. Lowe \cite{lowe:96,lowe:2}, found a flaw by
using an automatic tool FDR. The flaw was not detected in the
original proof because of different assumptions on the intruder
model. The fact that this new attack had escaped the attention of the experts was an indication of the underestimation of the
complexity of protocol analysis. This example has shown that protocol
analysis is critical for assessing the security of such cryptographic
protocols.

In this report, we present our approach for protocol
analysis together with a real example where we find an important flow
in a contemporary wireless sensor network security protocol. We start
by modelling protocols using a specific process algebraic formalism
called \LYSA process calculus. We then apply an analysis based on a
special program analysis technique called control flow analysis. We
apply this technique to the ZigBee-2007 End-to-End Application Key Establishment Protocol and with the help of the analysis discover
an unknown flaw. Finally we suggest a fix for the protocol, and verify
that the fix works by using the same technique.

\section{An Overview of the Analysis Method}
Static program analysis, in essence, examines a program statically,
before any attempt of execution. Although the finite amount of
resources may limit the information or the answers to important
questions, the approximation based approach of static program analysis makes
it preferable on the area of protocol analysis. Instead of facing
undecidability problem, this technique sacrifices precision and
gives approximate answers about a property of a certain program, or a
piece of code, or a protocol as in our case. However, this loss of
precision does not mean that we are missing the flaws, it merely
means that the analysis results may include false positives, such as a
bug or a flaw that the program does not contain.

Static program analysis was originally developed for generating codes
and optimising compilers \cite{Lowrey:Medlock,Busam:Englund}. Nevertheless, the analysis
technique have recently been directed to the field of
security. Encouraging results have been obtained by the use of this
approach where safe approximations to the set of values or behaviours
arising during protocol runs can be predicted.

Control flow analysis of processes formalised in the \LYSA process
calculus successfully computes an over-approximation of the run-time
behaviour of a protocol \cite{bod:2, bod:1}. This method is actually the
protocol analysis method that we present in this report.
The roadmap of the analysis method is given in Fig. \ref{fig:flow},
and we will present the steps of this roadmap in the following sections.
\begin{figure}[!t]
\centering
\includegraphics[width=3.5in]{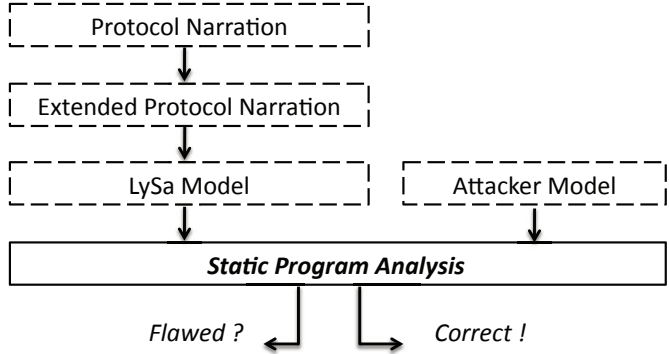}
\caption{The Roadmap of the Analysis}
\label{fig:flow}
\end{figure}

\section{Modelling in \LYSA Process Calculus}
\label{lys}
The first step in the protocol analysis is to formalise the protocol
narration into a model that is suitable for the analysis.
In our case, we formalize the protocols using the \LYSA process calculus \cite{bod:1}. 
\LYSA is based on the $\pi$-calculus \cite{mil} and incorporates cryptographic operations using ideas from the Spi-calculus \cite{aba:gor}. 
However, \LYSA has two different properties compared to spi/$\pi$ calculus. 
First, \LYSA has one global ether, instead of channels. 
The reason for this difference is that, in usual networking implementations (e.g. ethernet-based, wireless, etc.), 
anyone can eavesdrop or act as an active attacker which does not correspond to the channel-based communication. 
The second difference is in the pattern matching usage in the tests of the expressions associated with input and decryption.
Although \LYSA is a very powerful process calculus which also supports
asymmetric encryption, digital signatures, etc., in order to make it
simple we only illustrate the \emph{symmetric} fragment. 
The symmetric fragment suffices to prove our claims in the example
that we will present the flaw discovery since the protocol is designed
for symmetric encryption only.
The reader interested in further details including the asymmetric fragment may refer to \cite{bod:1}.
 
In \LYSA, we have terms ($E$) that consist of names (keys, nonces, messages, etc.), variables, and the compositions of them using symmetric encryption. 
The syntax of terms is shown in Table \ref{tab:terms}. 
In the case of encryption, the tuples of terms $E_1,\ldots,E_k$ are encrypted under a term $E_0$ which actually represents an encryption key. 
Note that an assumption of perfect cryptography is adopted, which means that \emph{decryption with the correct key} is the only inverse function of encryption. 
The \emph{annotation} inside brackets in the end of encryption will be explained later in this section.
 
\begin{table}
\caption{\LYSA Terms - Symmetric Fragment}
\label{tab:terms}
\centering
\begin{tabular}{rlll}
\hline
\multicolumn{2}{l}{$E$ ::=}  \\
              & $x$                                               &  $\mathrm{variable}$ \\
$\PAR$  & $n$                                               &  $\mathrm{name}$ \\
$\PAR$  & $\ANSENC{E_1,\ldots,E_k}{E_0}{\ell}{\mathcal{L}}$  &  $\mathrm{symmetric~encryption}$ \\
\hline
\end{tabular}
\end{table}
The syntax of the processes ($P$) which is mostly alike to the polyadic Spi-calculus \cite{aba:gor} is shown in Table \ref{tab:processes}. 
At this point, we prefer to skip the syntax for simple ones in the table, but explain the more interested and complicated two: output and input processes. 
The output process $\OUT{E_1,\ldots,E_k}.P$ sends the $k$-tuple $E_1,\ldots,E_k$ to the network and continues as process $P$.
Similarly, the input process $(E_1,\ldots,E_j ; x_{j+1},\ldots,x_k).P$ receives a $k$-tuple $E'_1,\ldots,E'_k$ and if conditions are satisfied, removes the $k$-tuple from the network.
Here, the input operation uses pattern matching which will only succeed if the prefix of the input message matches the terms specified before the semi-colon. 
In a simple manner, we can say that for some input $E'$ the input process ($E$;$x$).$P$ means that 
if $E'$ can be separated into two parts such that first part pairwise matches to the values $E$, then the remaining part of the input will be bound to the variables $x$. 
As you can see in Table \ref{tab:processes}, the number of tuples in $E'$ is $k$ so that this is the total number of tuples in $E$ and $x$. 
This kind of pattern matching is also used in decryption. 

\begin{table*}
\caption{\LYSA Processes - Symmetric Fragment}
\label{tab:processes}
\centering
\begin{tabular}{lll}
\hline
\multicolumn{2}{l}{$P$ ::=}  \\
             & $0$                                                             & $\mathrm{nil}$ \\
$\PAR$ & $P_1 \PAR P_2 $                                                 & $\mathrm{parallel~comp.}$ \\
$\PAR$ & !$P $                                                           & $\mathrm{replication}$\\
$\PAR$ & $\NEW{n} P $                                                    & $\mathrm{restriction}$\\
$\PAR$ & $\OUT{E_1,\ldots,E_k}.P $                                       & $\mathrm{output}$\\
$\PAR$ & $(E_1,\ldots,E_j ; x_{j+1},\ldots,x_k).P $                       & $\mathrm{input}$\\
$\PAR$ & $\ANSDEC{E}{E_1,\ldots,E_j}{x_{j+1},\ldots,x_k}{E_0}{\ell}{\mathcal{L}} P $    & $\mathrm{symm.~decryption}$\\
\hline
\end{tabular}
\end{table*}
\textbf{Example 1.a}  \textit{The example \LYSA code below is a \textbf{new} (created - restriction) encryption key ($K$) followed by an \textbf{output} which includes three plaintext elements ($A$, $B$, $K_A$) and an encrypted element ($\SENC{K}{K_A}{}{}$).}
\begin{center}$\NEW{K}\OUT{A,B, K_A, \SENC{K}{K_A}{}{} }$\end{center} 

\textbf{Example 1.b}  \textit{The example \LYSA code below is an \textbf{input} that binds the last two elements of the input to the variables $x_{KA}$ and $x$ as long as the first two elements are $A$ and $B$.}
\begin{center}$(A, B ; x_{KA}, x)$\end{center} 

\textbf{Example 1.c}  \textit{The example \LYSA code below is a \textbf{decryption} that decrypts the value bound to variable $x$ using the encryption key bound to variable $x_{KA}$ and binds the resulting plaintext value to the variable $x_K$. Note that this decryption always succeeds without any need of pattern matching, as long as the correct key exists in the receiver.}
\begin{center}$\SDEC{x}{}{x_K}{x_{KA}}{}$\end{center} 

In order to describe the \emph{message authentication} intentions of the protocols, we also have \emph{annotation}s for origin and destination.
Encryptions can be annotated with fixed labels called \textit{crypto-point}s that define their positions in the process, 
and with \textit{assertion}s that specify the origin and destination of encrypted messages.
A crypto-point \(\ell\) is an element of some set \(\mathcal{C}\) and used when encryptions/decryptions occur. 
The \LYSA term for encryption:
\begin{center} $\ANSENC{E_1,\ldots,E_k}{E_0}{\ell}{\mathcal{L}}$ \end{center}
means that the encryption happened at crypto-point \(\ell\) and the assertion [dest \(\mathcal{L}\)] means that 
corresponding (valid) decryption is \emph{to} happen at a crypto-point that belongs to the set $\mathcal{L}$ such that \(\mathcal{L} \subseteq \mathcal{C}\).
Similarly, in the \LYSA term for decryption:
\begin{center} $\ANSDEC{E}{E_1,\ldots,E_j}{x_{j+1},\ldots,x_k}{E_0}{\ell}{\mathcal{L}} P$ \end{center}
[orig \(\mathcal{L}\)] specifies the crypto-points \(\mathcal{L} \subseteq \mathcal{C}\) that $E$ is allowed to have been encrypted.

\textbf{Example 2}  \textit{The example \LYSA code below is the \textbf{composition} of the three separate parts in Example 1, and the necessary \textbf{annotations} in such a way that now we have two separate processes running in \textbf{parallel}.}

\begin{tabular}{ll}
         & \\
/* a */  & $\NEW{K}\OUT{A,B, K_A, \ANSENC{K}{K_A}{\ell_A}{\{\mathcal{\ell_B}\}} }.0$ \\
         & $\PAR$ \\
/* b */  & $(A, B ; x_{KA}, x).$ \\
/* c */  & $\ANSDEC{x}{}{x_K}{x_{KA}}{\ell_B}{\{\mathcal{\ell_A}\}}.0$ \\
         & \\
\end{tabular}

\textit{The example we constructed step by step is actually the \LYSA model of the single-message protocol below:}
\begin{center} \textbf{1. A \(\rightarrow\) B:} KA, \{K\}$_{KA}$ \end{center}
\textit{The upper part (line a) of the parallel composition is the code for principal A, and the lower part (lines b and c) is for principal B. In this example, annotations state that the encryption at crypto-point $\ell_A$ is intended to be decrypted only at $\ell_B$. In a corresponding manner, the decryption at $\ell_B$ should originate from the encryption at $\ell_A$.}

\subsection{Specifying Protocols in \LYSA}
In the beginning, we have a protocol narration like the one in Table \ref{tab:protocolnarration1}. 
Then we extend the narration to specify the internal actions to be performed in principals when receiving those messages. 
The reason for this kind of extension or conversion is to completely state the actions internal to the principals, which are normally left implicit in the narration of security protocols.

\begin{table}

\caption{Extended Protocol Narration - Case 1}
\label{tab:xpn}
\begin{tabular}{lllll}
\hline
\textbf{1. A} &\textbf{\(\rightarrow\)} &\textbf{ } &:& A, TC, \{TC, AppKey, B\}$_{KA}$ \\
               &                           &            & & [\textnormal{dest TC}] \\
\textbf{1'. } &\textbf{\(\rightarrow\)} &\textbf{TC}&:& $x_{initiator}$, $x_{TC}$, $x_{message}$ \\
               &                           &            & & [\textnormal{check } $x_{TC}$=TC]\\
\textbf{1''.} &                           &\textbf{TC}&:& \textnormal{decrypt $x_{message}$ as }\\
               &                           &            & & \{$x'_{TC}$, $x_{keytype}$, $x_{responder}$\}$_{KA}$\\
               &                           &            & & [\textnormal{orig }$x_A$][\textnormal{check } $x'_{TC}$=TC, $x_{keytype}$=AppKey]\\
               &                           &            & & \\
\textbf{2. TC}&\textbf{\(\rightarrow\)} &            &: & [\textnormal{new} LK]\\
               &                           &            && TC, $x_{initiator}$, \\
               &                           &            && \{$x_{initiator}$,AppLK,$x_{responder}$,TRUE,LK\}$_{KA}$ \\
               &                           &            && [\textnormal{dest }$x_{initiator}$] \\
\textbf{2'. } &\textbf{\(\rightarrow\)} &\textbf{A} &:& $y_{TC}$, $y_A$, $y_{message}$ \\
               &                           &            & & [\textnormal{check } $y_{TC}$=TC, $y_A$=A]\\
\textbf{2''.} &                           &\textbf{A} &:& \textnormal{decrypt $y_{message}$ as }   \\
               &                           &            & & \{$y'_A$, $y_{keytype}$, $y_B$, $y_{bool}$, $y_{LK}$\}$_{KA}$\\
               &                           &            & & [\textnormal{orig TC}][\textnormal{check } $y'_A$=A, $y_{keytype}$=AppLK]\\
               &                           &            & & [\textnormal{check } $y_B$=B, $y_{bool}$=TRUE] \\
\hline
\end{tabular}
\end{table}
As an example, the extended protocol narration of (due to the lack of space) the first two messages of Case 1 is given in Table \ref{tab:xpn}. 
For each message in the original protocol narration, we have an output message $n$ and an input message $n'$ in the extended protocol narration.
Input message $n'$ presents the variable (those written in \emph{italics}) bindings and necessary checks in the receiver side.
If a variable is a ciphertext and the receiver has the correct encryption key, then we have another message (i.e. $n''$) for each of those variables.
In addition, we explicitly write the internal actions as annotations between square brackets, in order to bridge the gap between informal and formal specification of the protocol.
Note that when analysing protocols we add an extra message to the end, where a principal attempts to communicate the other through the new shared key, LK. For example, the message
\begin{center} \textbf{1. B \(\rightarrow\) A:} \{MSG\}$_{LK}$ \end{center}
does not change the protocol nor bring any (nor bring any additional cost to the implementations), it is just a sample message that will be sent using the new LK and 
thus it will ease the validation which is done by checking attackers knowledge.

In the next phase, we convert the extended protocol narration into a \LYSA model. 
We use the \LYSA syntax that we explained earlier in this section and configure the necessary settings.
As an example, a regular \LYSA model of the protocol that we have used
to demonstrate extended protocol conversion is given in Table \ref{tab:lysaC1}.
Further details of specifying protocols in \LYSA are present in \cite{bod:1}.

\begin{table}
\caption{LySa Model - Case 1}
\label{tab:lysaC1}
\centering
\begin{tabular}{l@{}l}
\hline
             & $\LET{X}{\mathbf{N}\;\LFONT{s.t.}\;\CANON{\mathbf{N}} = \{1,2,3\}}\IN$\\
             & $\INEW{i\in X}{\mathit{KA}_{i}}$ $\INEW{j\in X}{\mathit{KB}_{j}}$\\
             
& $\IPAR{i\in X}\IPAR{j\in X\cup \{ 0 \}} !$\\
\textbf{1}   & $\OUT{A_{i}, \mathit{TC}, \SENC{\mathit{TC}, \mathit{AppKey}, B_{j}}{\mathit{KA}_{i}}\ATDEST{\mathit{a1}_{i j}}{\{\mathit{tc1}_{i j}\}}}.$\\
\textbf{2'}  & $\INP{\mathit{TC}, A_{i}}{\mathit{y}_{i j}}.$\\
\textbf{2''} & $\SDEC{\mathit{y}_{i j}}{A_{i}, \mathit{AppLK}, B_{j}, \mathit{TRUE}}{\mathit{xLK}_{i j}}{\mathit{KA}_{i}}$\\
              & $\ATORIG{\mathit{a2}_{i j}}{\{\mathit{tc2}_{i j}\}}\IN$\\
\textbf{4'}  & $\INP{B_{j}, A_{i}}{\mathit{y2}_{i j}}.$\\
\textbf{4'' } & $\SDEC{\mathit{y2}_{i j}}{}{\mathit{xmsg}_{i j}}{\mathit{xLK}_{i j}}\ATORIG{\mathit{a4}_{i j}}{\{\mathit{b4}_{i j}\}}\IN$ $\NIL$\\
              & $\PAR$\\
              & $\IPAR{j\in X}\IPAR{i\in X\cup \{ 0 \}} !$\\
\textbf{3'}  & $\INP{\mathit{TC}, B_{j}}{\mathit{z}_{i j}}.$\\
\textbf{3''} & $\SDEC{\mathit{z}_{i j}}{B_{j}, \mathit{AppLK}, A_{i}, \mathit{FALSE}}{\mathit{yLK}_{i j}}{\mathit{KB}_{j}}$\\
              & $\ATORIG{\mathit{b3}_{i j}}{\{\mathit{tc3}_{i j}\}}\IN$\\
\textbf{4}   & $\NEW{\mathit{MSG}_{i j}} \OUT{B_{j}, A_{i}, \SENC{\mathit{MSG}_{i j}}{\mathit{yLK}_{i j}}\ATDEST{\mathit{b4}_{i j}}{\{\mathit{a4}_{i j}\}}}.$ $\NIL$\\
              & $\PAR$\\
              & $\IPAR{i\in X\cup \{ 0 \}}\IPAR{j\in X\cup \{ 0 \}} !$\\
\textbf{1'}  & $\INP{A_{i}, \mathit{TC}}{\mathit{x}_{i j}}.$\\
\textbf{1''} & $\SDEC{\mathit{x}_{i j}}{\mathit{TC}, \mathit{AppKey}, B_{j}}{}{\mathit{KA}_{i}}$\\
              & $\ATORIG{\mathit{tc1}_{i j}}{\{\mathit{a1}_{i j}\}}\IN$\\
\textbf{2}   & $\NEW{\mathit{LK}_{i j}} \langle\mathit{TC}, A_{i}, \SENC{A_{i}, \mathit{AppLK}, B_{j}, \mathit{TRUE}, \mathit{LK}_{i j}}{\mathit{KA}_{i}}$\\
              & $\ATDEST{\mathit{tc2}_{i j}}{\{\mathit{a2}_{i j}\}}\rangle.$\\
\textbf{3}   & $\langle\mathit{TC}, B_{j}, \SENC{B_{j}, \mathit{AppLK}, A_{i}, \mathit{FALSE}, \mathit{LK}_{i j}}{\mathit{KB}_{j}}$\\
              & $\ATDEST{\mathit{tc3}_{i j}}{\{\mathit{b3}_{i j}\}}\rangle.$ $\NIL$\\
\hline
\end{tabular}
\end{table}

\section{Static Program Analysis}
\label{zig:static}
Static Analysis is a formal method that enables the security analysis of cryptographic communication protocols which are modelled as \LYSA processes. 
Messages communicated on the network are tracked with the possible values of the variables in the protocol. 
Besides, the potential violations of the destination/origin annotations are also recorded.
The aim of static analysis is to efficiently compute the safe approximations to the behaviour of the models without actually running them.
In Fig. \ref{fig:stat} we can see the approximation approach.
In general, it is impossible to compute the precise answer so we make a choice between over-approximation and under-approximation.
Static analysis over-approximates the set of possible operations that the \LYSA process describes.
The nature of over-approximation may cause the analysis to investigate a trace which is impossible at all.  
However, over-approximation is needed to make a safe approximation since under-approximation could miss some traces.

\begin{figure}[!htp]
\centering
\includegraphics[width=3.5in]{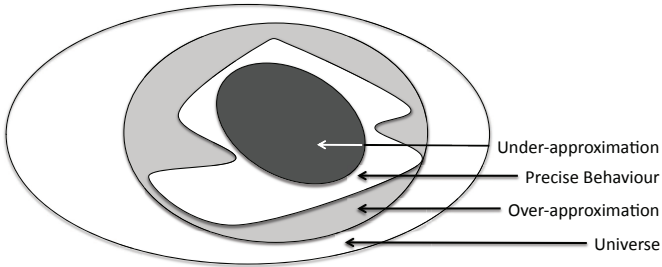}
\caption{Static Analysis}
\label{fig:stat}
\end{figure}

\subsection{Analysis Method}

The static analysis we use in this study is specified as a Flow Logic
\cite{bod:2,bod:1}, which is based on the control flow analysis and
the data flow analysis techniques that allow us to make it fully
automatic \cite{nie:nie:han}. 

Control flow analysis is a program analysis technique that is used to
compute approximations of the result of a program execution without
running the program. Such an analysis helps us in determining the sets
of values that may be generated by communication using a specific
protocol, which is beneficial for validating certain security
properties. Especially when used in conjunction with a model of
possible malicious activity (i.e. attacker), the analysis provides a
safe approximation of all events that may happen.

Flow Logic is a notational style for specifying analyses across
programming paradigms, introduced by Nielson, Nielson 
\cite{flow:1,flow:2,flow:3}, and with Hankin
\cite{nie:nie:han}. 
By abstracting from domain
specific formalisms and instead using standard mathematical notations, the Flow
Logic constitutes a meta-language that can present an analysis without requiring
additional knowledge about particular formalisms. Deriving an analysis estimate
from the resulting analysis specification is then left as a separate activity, usually
involving orthogonal considerations and tools.
This approach allows the designer to focus on the specification of analyses without
making compromises dictated by implementation considerations. Similarly,
implementation is simplified and improved, as the implementer is always free to
choose the best available tool. In the next sections, we will present
control flow analysis of \LYSA in the style of flow logic.

The control flow analysis that we use in protocol analysis is specified using the flow logic framework as a predicate
\begin{center}$\JUDGE{\rho, \kappa, \psi}{P}$\end{center}
that holds precisely when $\rho $, $\kappa$, and  $\psi $ form an analysis result that correctly
describes the behaviour of the process $P$. 

The main components of the analysis are:
\begin{itemize}
\item \emph{The variable environment $\rho$,} an over-approximation of the potential values of each variable that it may be bound to.
\item \emph{The network component $\kappa$,} an over-approximation of the set of messages that can be communicated over the network.
\item \emph{The error component $\psi$,} the set of error messages in the form $(\ell,\ell')$, indicating that something encrypted at $\ell$ was unexpectedly decrypted at  $\ell'$.
\end{itemize}


The analysis is judgments of the form  $\JUDGE{\rho, \kappa, \psi}{P}$ which express that $\rho, \kappa, \psi$ compose a valid analysis for the process $P$. We also need to introduce the auxiliary judgment $\JUDGE{\rho}{E \LFONT{ : } \vartheta}$ at this point. This expresses that $\vartheta$, the set of values, is an acceptable estimate of the values that the term $E$ may evaluate in $\rho$, the abstract environment.

\label{lys:can}To keep the analysis component finite, we partition all the names that are generated by a \LYSA process into finitely many equivalence classes.
A \emph{canonical value} is a representative for each of these equivalence classes. 
Names from the same equivalence class are assigned a common \emph{canonical name} and instead of the actual names, we use the names of those equivalence classes.
For example, the canonical representative of a name $n$ is denoted by $\lfloor n \rfloor$.
Since it allows us to analyse an infinite number of principals, canonical value is an important analysis element \cite{buc:nie}.

The analysis of terms is listed in Table \ref{tab:analysisterm}. 
The rule for analysing names $\LFONT{(AName)}$ states that $\vartheta$ is an acceptable estimate for a name $n$ if the canonical representative of $n$ belongs to $\vartheta$. 
The rule for analysing variables $\LFONT{(AVar)}$ states that $\vartheta$ is an acceptable estimate for a variable $x$ if it is a superset of  $\rho (\lfloor x \rfloor)$.
The rule for analysing symmetric encryption $\LFONT{(AEnc)}$ finds the set $\vartheta_i$ for each term $E_i$, collects all k-tuples of values $(V_0,\ldots,V_k)$ taken from $\vartheta_0 \times \ldots \times \vartheta_k$ into values of the form $\ANSENC{V_1,\ldots,V_k}{V_0}{\ell}{\mathcal{L}}$ and requires that these values belong to $\vartheta$.

\begin{table*}
  \caption{Analysis for Terms, $\JUDGE{\rho}{E \LFONT{ : } \vartheta}$} 
 \label{tab:analysisterm}
\centering
   \begin{tabular}{lc}
\hline
     \LFONT{(AName)}      & \INFERENCE{\lfloor n \rfloor \in \vartheta}{\JUDGE{\rho}{n \LFONT{ : } \vartheta}}  \\
 & \\
     \LFONT{(AVar)}   & \INFERENCE{\rho (\lfloor x \rfloor) \subseteq \vartheta}{\JUDGE{\rho}{x \LFONT{ : } \vartheta}}  \\
 & \\
\multirow{2}{*}{\LFONT{(AEnc)}}  & $\JUDGE{\wedge_{i=0}^k \rho}{E_i \LFONT{ : } \vartheta_i} \quad \wedge$ \\
 & \INFERENCE{\forall V_0, V_1,\ldots,V_k \LFONT{ : } \wedge_{i=0}^k V_i \in \vartheta_i \quad  \Rightarrow \quad \ANSENC{V_1,\ldots,V_k}{V_0}{\ell}{\mathcal{L}} \in \vartheta}{\JUDGE{\rho}{\ANSENC{E_1,\ldots,E_k}{E_0}{\ell}{\mathcal{L}} \LFONT{ : }\vartheta}}  \\
\hline
   \end{tabular}
\end{table*}

The analysis of processes is listed in Table \ref{tab:analysisproce}. The idea of the analysis is very similar to the analysis of terms, therefore instead of explaining all the rules we explain only one interesting rule.
The rule for analysing output $\LFONT{(AOut)}$ uses the analysis for terms to find the estimate  $\vartheta_i$ for each term $E_i$ and requires that all all k-tuples of values $\OUT{V_1,\ldots,V_k {}{} }$ taken from $\vartheta_1 \times \ldots \times \vartheta_k$ are in $\kappa$ (i.e. they may flow on the network). The rule also requires that the components $\rho, \kappa, \psi$ compose a valid analysis for process $P$.

\begin{table*}
  \caption{Analysis for Processes, $\JUDGE{(\rho , \kappa)}{P \LFONT{ : } \psi}$} 
 \label{tab:analysisproce}
\centering
   \begin{tabular}{lc}
\hline
     \LFONT{(ANil)}   & \JUDGE{(\rho , \kappa)}{0 \LFONT{ : } \psi}  \\
 & \\
     \LFONT{(APar)}   & \INFERENCE{\JUDGE{(\rho , \kappa)}{P_1 \LFONT{ : } \psi} \quad \wedge \quad \JUDGE{(\rho , \kappa)}{P_2 \LFONT{ : } \psi}}{\JUDGE{(\rho , \kappa)}{P_1 \PAR P_2 \LFONT{ : } \psi}}  \\
 & \\
      \LFONT{(ARep)}  & \INFERENCE{\JUDGE{(\rho , \kappa)}{P \LFONT{ : } \psi}}{\JUDGE{(\rho , \kappa)}{\LFONT{!} P \LFONT{ : } \psi}}  \\
 & \\
      \LFONT{(ANew)}  & \INFERENCE{\JUDGE{(\rho , \kappa)}{P \LFONT{ : } \psi}}{\JUDGE{(\rho , \kappa)}{\NEW{n} P \LFONT{ : } \psi}}  \\
 & \\
\multirow{3}{*}{\LFONT{(AOut)}}  & $\JUDGE{\wedge_{i=1}^k \rho}{E_i \LFONT{ : } \vartheta_i} \quad \wedge $ \\
   & $\JUDGE{(\rho , \kappa)}{P \LFONT{ : } \psi}$\\
   & $\INFERENCE{\forall V_1,\ldots,V_k \LFONT{ : } \wedge_{i=1}^k V_i \in \vartheta_i \quad  \Rightarrow \quad \OUT{V_1,\ldots,V_k {}{} } \in \kappa \quad \wedge \quad}{\JUDGE{(\rho,\kappa)}{\OUT{E_1,\ldots,E_k {}{} }. P \LFONT{ : } \psi}}$ \\
 & \\
\multirow{3}{*}{\LFONT{(AIn)}}  & $\JUDGE{\wedge_{j=1}^k \rho}{E_i \LFONT{ : } \vartheta_i} \quad \wedge \quad$ \\
    & $\JUDGE{(\rho , \kappa)}{P \LFONT{ : } \psi}$\\
    & $\INFERENCE{\forall V_1,\ldots,V_k \in \kappa \LFONT{ : } \wedge_{j=1}^k V_i \in \vartheta_i \quad  \Rightarrow \quad \wedge_{i=j+1}^k V_i \in \rho (\lfloor x_i \rfloor) \quad \wedge \quad}{\JUDGE{(\rho , \kappa)}{\INP{E_1,\ldots,E_j}{x_{j+1},\ldots,x_k}. P \LFONT{ : } \psi}}$  \\
 & \\
\multirow{5}{*}{\LFONT{(ADec)}}  & $\JUDGE{\rho}{E \LFONT{ : } \vartheta} \quad \wedge \quad$ \\
    & $\JUDGE{\forall \wedge_{i=0}^j \rho}{E_i \LFONT{ : } \vartheta_i} \quad  \wedge \quad$ \\
    & $\JUDGE{(\rho , \kappa)}{P \LFONT{ : } \psi}$ \\
    & $((\ell \notin \mathcal{L}' \vee \ell ' \notin \mathcal{L}) \Rightarrow (\ell , \ell ') \in \psi) \quad \wedge \quad$ \\
    & $\INFERENCE{\forall \ANSENC{V_1,\ldots,V_k}{V_0}{\ell}{\mathcal{L}} \in \vartheta \LFONT{ : } \wedge_{i=0}^j V_i \in \vartheta_i \quad  \Rightarrow \quad \wedge_{i=j+1}^k V_i \in \rho (\lfloor x_i \rfloor) \quad \wedge \quad}{\JUDGE{(\rho , \kappa)}{\ANSDEC{E}{E_1,\ldots,E_j}{x_{j+1},\ldots,x_k}{E_0}{\ell '}{\mathcal{L}} P \LFONT{ : } \psi}}$ \\
\hline
   \end{tabular}
\end{table*}

\textbf{Example 3}  \textit{Static analysis of the \LYSA model given in Example 2 will lead to the following results:}\\
\begin{tabular}{ll}
 & \\
 & $\OUT{A,B, K_A, \ANSENC{K}{K_A}{\ell_A}{\mathcal{\ell_B}} } \in \kappa$ \\
 & $K_A \in \rho(x_{KA})$ \\
 & $\ANSENC{K}{K_A}{\ell_A}{\mathcal{\ell_B}} \in \rho(x)$ \\
 & $K \in \rho(x_K)$ \\
 & \\
\end{tabular}

\textit{Looking at the results above, it is easy to see that the first line is related to \textbf{line a} in Example 2. Likewise, next two lines derived from \textbf{line b} and the last line derived from \textbf{line c} in Example 2. Note that, \textit{how the analysis works} is not the subject of this paper. Therefore, see \cite{bod:1} for \textit{how} Example 2 leads to Example 3.}

\subsection{Attacker Model}
In practice, network protocols are vulnerable to attacks. 
Unfortunately it is even easier to attack wireless networks since any computer within range that is equipped with a wireless client card can pull the signal and access the data.
In this study, \LYSA processes are analysed in parallel with the Dolev-Yao attacker \cite{dol:yao}. 
The operations that this attacker model can perform are listed below, 
but before this we have to introduce new canonical (see Section~\ref{lys:can}) names and variables for the attacker.
All the canonical names of the attacker are mapped to $n_\bullet$ and all the canonical variables of the attacker are mapped to $z_\bullet$. 
We also have \(\ell\)$_\bullet$ which is a crypto-point in the attacker.

The descriptions of the Dolev-Yao conditions are:
\begin{itemize}
  \item The attacker initially has the knowledge of the canonical name $n_\bullet$ and all free names of the process $P$ but he can improve his knowledge by eavesdropping on all messages sent on the network.
  \item The attacker can improve his knowledge by decrypting messages with the keys he already knows. Unless the intended recipient of the message was an attacker, an error (\(\ell\),\(\ell\)$_\bullet$) should be added to the error component $\psi$ which means that something encrypted at \(\ell\) was actually decrypted by the attacker at \(\ell\)$_\bullet$.
  \item The attacker can construct new encryptions using the keys he already knows. If this message is received and decrypted by a principal, then an error (\(\ell\)$_\bullet$,\(\ell\)) should be added to the error component $\psi$
       which means that something encrypted at the attacker was decrypted by the attacker by a process $P$ at \(\ell\).
  \item The attacker can send messages on the network using his knowledge and thus forge new  communications.
\end{itemize}

These conditions enable the attacker to establish scenarios including eavesdropping, modification, man-in-the-middle and replay attacks. 
The soundness of the Dolev-Yao condition is proved in \cite{bod:1}.

As shown in Fig. \ref{fig:flow}, the \LYSA model of a protocol is analysed in parallel with the attacker model and processed by the \LYSA-tool (see Section \ref{zig:formal}) which implements the static analysis. 
The results of the analysis are used to validate destination/origin authentication and confidentiality properties of the protocols. 
If no violation is detected, namely the error component $\psi$ is empty, then it is guaranteed that the protocol satisfies the destination/origin authentication properties. 
Furthermore, the potential values that are learned by the attacker help us in validating the confidentiality properties. 
The details as well as the proof of the soundness of the analysis are presented in \cite{bod:2}.

\textbf{Example 4}  \textit{In Example 3, we analysed Example 2 in an attack-free setting. Now we add the attacker model and get the following results in addition to the results in Example 3. Since the attacker is able to learn everything sent on the network we have:}
\begin{center}$K_A,\ANSENC{K}{K_A}{\ell_A}{\mathcal{\ell_B}} \in \rho(z_\bullet)$ \end{center}
\textit{Therefore, the attacker can decrypt the encrypted part of the message which leads to the violation:}
\begin{center}($\ell_A, \ell_\bullet$) $\in \psi$\end{center}
\textit{Thus we conclude that the encryption at crypto-point $\ell_A$ which was intended to be decrypted at $\ell_B$ can be decrypted by the attacker and hence the example protocol is flawed.}
\section{Application on ZigBee Wireless Sensor networks}
In this section, we present an application of the analysis method that
we explained up to now \cite{yuk:nie:2009}. This application has many features that make
it interesting. First of all, it pinpoints an undiscovered and
non-trivial flaw in a
real cryptographic security protocol. Another key issue is that the
protocol is being used in one of the latest wireless sensor network
standards, ZigBee, that is promising and emerging in the sensor
networks field. Therefore, the protocol includes secure components
that are known to be secure when they are individually used and some
of them are industry standards such as SKKE that we will explain in
more details. Still we show that combining proven to be secure
components is not sufficient for guaranteeing security
properties. Last feature of this application is that we not only use
protocol analysis to discover flaws but also to verify our fix
proposals. 

\subsection{ZigBee-2007 End-to-End Application Key Establishment Protocol}
\label{zig:keyest}
ZigBee is a fairly new but promising Wireless Personal Area network (WPAN) standard for wireless sensor
networks
that have very low resource requirements. In parallel with this, the devices that operate in
ZigBee networks have limited resources in terms of memory, processor,
storage, power, etc. Therefore
implementing the security guarantees is a great challenge and the
verification of the security properties is of paramount importance.

We start by presenting the key points that are necessary for a clear understanding of the development, and we omit all the details which are not directly relevant to this study. However, a detailed survey on ZigBee security can be suggested as \cite{Yuksel:Nielson} and surely the ultimate source is the ZigBee documentation \cite{ZigBee:2007, ZigBee:Stack, ZigBeePRO:Stack, ZigBee:HA, ZigBee:SE} which is a rather difficult read with hundreds of pages including references to several other standards.

\begin{figure}[!htp]
\centering
\includegraphics[width=3.3in]{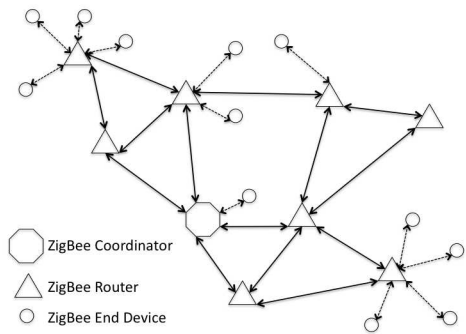}
\caption{ZigBee Network Model}
\label{fig:znet}
\end{figure}
End-to-End Application Key Establishment is the protocol to be used when establishing a Link Key (\textbf{LK}) between two ZigBee devices, 
which are running in \emph{High Security Mode} (which was called \emph{Commercial Mode} in the previous standard, ZigBee-2006 \cite{ZigBee:2006}). 
We will call the devices as initiator (\textbf{A}) and responder (\textbf{B}). 
Note that there is also a Trust Center (\textbf{TC}), which shares a pairwise secret key with each principal in the network. 
TC is actually an application that runs on a preferably more powerful ZigBee device referred to as \emph{ZigBee Coordinator} which is unique in the network; 
whereas the remaining devices might be of type \emph{ZigBee Router} or \emph{ZigBee End Device}, as shown in Fig. \ref{fig:znet}.
For a better understanding we should mention that for two ZigBee devices to establish a secure communication, 
they must share a symmetric key (LK) which they either \emph{receive} from a trusted server (TC) 
or \emph{create mutually} using a temporary key received from the trusted server.

\begin{figure}[!htp]
\centering
\includegraphics[width=3.3in]{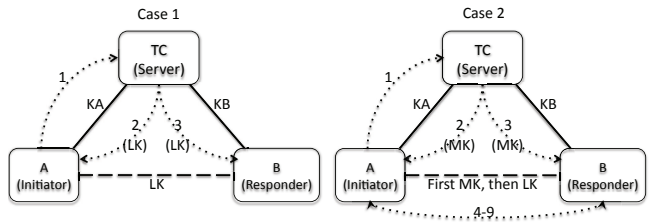}
\caption{ZigBee-2007 End-to-End Application Key Establishment Scenarios}
\label{fig:scena}
\end{figure}
The scenarios of End-to-End Application Key Establishment are visualized in Fig. \ref{fig:scena}. 
The solid lines represent the already secure communication paths, labeled by corresponding symmetric encryption keys.
The dashed lines represent the resulting secure communication paths after a successful protocol run, again labeled by corresponding encryption keys.
Finally, the dotted lines are the messages in the protocol labeled by their sequence numbers and the encryption keys they deliver.

ZigBee-2007 End-to-End Application Key Establishment Protocol has two different cases according to the configuration of TC, we will call them as \emph{Case 1} and \emph{Case 2}. 
In Case 1, TC creates the LK itself and sends it to each principal. 
Therefore, the initiator and the responder have no role in the creation of the LK. 
In Case 2, TC creates a temporary shared key called \emph{Master Key} (\textbf{MK}) and sends it to each principal.
Using this MK, A and B initiate a Symmetric-Key Key Establishment (\textbf{SKKE}) procedure to establish an LK. 
This case allows principals to create an LK \emph{mutually}. 
SKKE is actually a key agreement scheme employed in the ZigBee End-to-End Application Key Establishment mechanism, and its components are defined in the ANSI X9.63-2001 standard \cite{ansi:x963}.
At the end of (a successful run of) either case, two ZigBee devices will be able to establish secure communication using their pairwise encryption key, LK.

\subsubsection{Case 1}

In Case 1, the initiator begins the procedure of establishing an LK with the responder by sending TC the first message, \textbf{request key},
which includes \emph{destination address} (\begin{footnotesize}=TC\end{footnotesize}), \emph{requested key type} (\begin{footnotesize}=Application Key\end{footnotesize}), and \emph{partner address} (\begin{footnotesize}=B\end{footnotesize}). 
Then TC creates an LK for two principals, and sends it to each principal in two similar \textbf{transport key} messages. 
Since TC is configured to send an LK directly in this case, the \emph{key type} value in the last two messages will be Application Link Key (AppLK). 
The only difference between these two messages is a boolean value that indicates the initiator 
(\begin{footnotesize}TRUE\end{footnotesize}: message recipient is the initiator, \begin{footnotesize}FALSE\end{footnotesize}: message recipient is the responder), and also the principal address'. 
All the messages in this case are encrypted with the sender/receiver principal's key that is shared with TC (assuming that the security suite is \emph{Encryption-only}). 
The type of this key can either be Trust Center Link Key (\textbf{TCLK}) or Trust Center Master Key (\textbf{TCMK}), as defined in the ZigBee specification \cite{ZigBee:2007}, 
but for simplicity we will call it \emph{KA} for principal A, and \emph{KB} for principal B.
The protocol narration of Case 1 is given in Table \ref{tab:protocolnarration1}.

\begin{table}
\caption{Protocol Narration - Case 1}
\label{tab:protocolnarration1}
\centering
\begin{tabular}{l}  
\hline
\textbf{1. A\(\rightarrow\)TC:} \{TC, AppKey, B\}$_{KA}$   \\
\textbf{2. TC\(\rightarrow\)A:} \{A, AppLK, B, TRUE, LK\}$_{KA}$  \\
\textbf{3. TC\(\rightarrow\)B:} \{B, AppLK, A, FALSE, LK\}$_{KB}$  \\
\hline
\end{tabular}
\end{table}

\subsubsection{Case 2}

In Case 2, the first three messages are almost the same as in Case 1, 
except in this case TC is configured to send MK, and therefore key type is the Application Master Key (AppMK). 
The rest of the messages are between the initiator and the responder. 
In the fourth message, \textbf{establish key}, A sends B his request to start SKKE. 
The values, False and Zero, indicate that there is no \emph{parent} (router, TC, etc.), and no \emph{parent address}, respectively. 
The fifth message is the response of B to A's SKKE request. 
Note that these two messages are encrypted by MK, which was received in the previous two messages. 
The remaining four messages are actually the SKKE protocol itself.
Messages 6 and 7 include the \emph{challenges} (NA, NB) of the principals. 
Messages 8 and 9 are the complex messages which can be computed by both parties to verify each other.
A and B create two \emph{message authentication codes} (\textbf{MAC}) 
using their knowledge, besides the MAC key itself is a \emph{hash} (H) of another MAC which they produce using the same knowledge \cite{HMAC}. 
After the verification, the new LK will be H(MAC\{A,B,NA,NB\}$_{MK}$, 2), which is a minor variation of the MAC key that was used in the last two messages.
The protocol narration of Case 2 is given in Table \ref{tab:protocolnarration2}.

\begin{table}
\caption{Protocol Narration - Case 2}
\label{tab:protocolnarration2}
\centering
\begin{tabular}{l}  
\hline
\textbf{1. A\(\rightarrow\)TC:} \{TC, AppKey, B\}$_{KA}$   \\
\textbf{2. TC\(\rightarrow\)A:} \{A, AppMK, B, TRUE, MK\}$_{KA}$ \\
\textbf{3. TC\(\rightarrow\)B:} \{B, AppMK, A, FALSE, MK\}$_{KB}$ \\
\textbf{4. A\(\rightarrow\)B:} \{B, FALSE, Zero, SKKE\}$_{MK}$ \\
\textbf{5. B\(\rightarrow\)A:} \{A, TRUE\}$_{MK}$                 \\                           
\textbf{6. A\(\rightarrow\)B:} \{NA\}$_{MK}$      \\                
\textbf{7. B\(\rightarrow\)A:} \{NB\}$_{MK}$                  \\                       
\textbf{8. A\(\rightarrow\)B:} MAC\{3,A,B,NA,NB\}$_{H(MAC\{A,B,NA,NB\}_{MK},1)}$        \\
\textbf{9. B\(\rightarrow\)A:} MAC\{2,B,A,NB,NA\}$_{H(MAC\{A,B,NA,NB\}_{MK},1)}$        \\
\hline
\end{tabular}
\end{table}

\subsection{The Flaw}
\label{flaw}
In wireless networks, it is easy to intercept, forge and inject messages. 
Without any formal analysis, an experienced eye can see that all the messages in ZigBee-2007 End-to-End Application Key Establishment Protocol can be replayed 
when the same long-term encryption keys (KA, KB) are still being used. 
The reason is the lack of \emph{freshness} elements like nonces, timestamps, etc. 
This flaw can lead to serious replay attacks, denial of service (DoS) attacks, etc. 
Even worse, when an old session key is compromised, an attacker can decrypt all the messages by replaying that old session key. In other words, lack of freshness can cause failures in \emph{authenticity} (in the case that principals accept an old session key from a rogue TC) and \emph{confidentiality} (in the case that principals start using a compromised session key).

As can be seen in the narration of the protocol, no freshness indicator is used in the distribution of either LK (in Case 1) or MK (in Case 2, the first three messages). 
Therefore, all the messages can be replayed. 
Replay of a message that includes a key is very critical. 
An attacker can store a message including a key from a previous run of this protocol, and then send the old message to make principals communicate using this old key. 
If the old key is compromised, then the attacker will be able to decrypt all the messages between two victim principals. 

The significance of the security risk that is caused by this flaw may require more explanation. 
Indeed, the flaw does not disclose any session key but allow reuse of a former key. 
Besides, brute force attacks or other types of known cryptographic
attacks for obtaining the key do not seem practical for the current
specification (i.e. the keys are 128-bits).
However, disclosure of a key might still be possible without dealing with cryptography, and reuse of an old session key can cause serious risks.
An example scenario is given below:

\textbf{Scenario 1}  \textit{A and B established a link key, and had secure communication with the help of that pairwise key. 
Than B left the network and disclosed the key, which might be by means
of hardware (e.g. local key extraction from the chipset such as  connecting a debugger, erasing the chip, then freely reading the contents of RAM), 
or software (e.g. a bug in the implementation that discloses the key
after the session expires or terminates with the natural assumption
that a new session key will be used for a future session) defects. If
B rejoins the network, and run the key establishment protocol with A (no
matter which case or security level is chosen), the disclosed key may
be replayed by the attacker who can decrypt all the communication using
the disclosed key.}

In the ZigBee Specification, the notion of \emph{frame counter} is emphasized as the freshness protection. 
This approach is not a strong one for several reasons.
First of all, a frame counter uses incrementing values rather than random values and rejects frames with a smaller counter value.
Second, regardless of the length (which is 32-bits in ZigBee) it is easy to cause overflow to frame counters. As indicated in \cite{Sastry:Wagner}, if an adversary forges a frame with the maximum value (\emph{i.e. 0xFFFFFFFF}) any further frame will be rejected.
In addition, using counters is not a novel approach, since in such layered architectures lower layers also used similar counters.

\subsubsection{Flaw in Case 1}
The attack scenario for Case 1 is given in Table \ref{tab:att1}. The first run (messages 1 to 3), is an old run which is intercepted by an attacker. Here, it is appropriate to mention that LK is used like a session key and KA/KB are used like master keys. Therefore, KA and KB are possibly the same in two different runs. The second run in the attack scenario (messages 1' to 3') is initiated regularly, but the last two messages are replayed by the attacker using the messages that are captured from the old run.
Furthermore, the attacker does not necessarily need to wait for a message like 1' since he can already replay it, too.

\begin{table}
\caption{Attack Scenario - Case 1}
\label{tab:att1}
\centering
\begin{tabular}{l}
\hline
\textbf{1. A\(\rightarrow\)TC:}      \{TC, AppKey, B\}$_{KA}$   \\
\textbf{2. TC\(\rightarrow\)A:}      \{A, AppLK, B, TRUE, LK\}$_{KA}$  \\
\textbf{3. TC\(\rightarrow\)B:}      \{B, AppLK, A, FALSE, LK\}$_{KB}$  \\
\hline
\textbf{1'. A\(\rightarrow\)TC:}     \{TC, AppKey, B\}$_{KA}$   \\
\textbf{2'. M(TC)\(\rightarrow\)A:}  \{A, AppLK, B, TRUE, LK\}$_{KA}$  \\
\textbf{3'. M(TC)\(\rightarrow\)B:}  \{B, AppLK, A, FALSE, LK\}$_{KB}$  \\
\hline
\end{tabular}
\end{table}

\subsubsection{Flaw in Case 2}
The attack for Case 1 is also possible for Case 2, in which MK is sent without any freshness indicator. 
Even though LK is created mutually by the use of SKKE in Case 2, 
a compromised old MK that is replayed to principals before SKKE will allow an attacker to create the LK as well. 
The attack scenario for Case 2 is given in Table \ref{tab:att2}. 
The first run (messages 1 to 9) is the old run and it is sufficient for an attacker to capture messages 2 and 3. 
Then the attacker replays these messages in the new run (messages 1' to 9').
Although the nonce's used in SKKE (exchanged in messages 6 and 7) are different, 
as long as MK is compromised the attacker can decrypt these messages and learn the nonces as well.
As a result, the attacker can still compute the new LK which is actually H(MAC\{A,B,NA',NB'\}$_{MK}$, 2) (see Section \ref{zig:keyest}).
Therefore, we may conclude that the flaw is critical in both cases. 

\begin{table}
\caption{Attack Scenario - Case 2}
\label{tab:att2}
\centering
\begin{tabular}{l}
\hline
\textbf{1. A\(\rightarrow\)TC:} \{TC, AppKey, B\}$_{KA}$   \\
\textbf{2. TC\(\rightarrow\)A:} \{A, AppMK, B, TRUE, MK\}$_{KA}$ \\
\textbf{3. TC\(\rightarrow\)B:} \{B, AppMK, A, FALSE, MK\}$_{KB}$ \\
\textbf{4. A\(\rightarrow\)B:} \{B, FALSE, Zero, SKKE\}$_{MK}$ \\
\textbf{5. B\(\rightarrow\)A:} \{A, TRUE\}$_{MK}$                 \\                           
\textbf{6. A\(\rightarrow\)B:} \{NA\}$_{MK}$      \\                
\textbf{7. B\(\rightarrow\)A:} \{NB\}$_{MK}$                  \\                       
\textbf{8. A\(\rightarrow\)B:} MAC\{3,A,B,NA,NB\}$_{H(MAC\{A,B,NA,NB\}_{MK},1)}$        \\
\textbf{9. B\(\rightarrow\)A:} MAC\{2,B,A,NB,NA\}$_{H(MAC\{A,B,NA,NB\}_{MK},1)}$        \\
\hline
\textbf{1'. A\(\rightarrow\)TC:} \{TC, AppKey, B\}$_{KA}$   \\
\textbf{2'. M(TC)\(\rightarrow\)A:} \{A, AppMK, B, TRUE, MK\}$_{KA}$ \\
\textbf{3'. M(TC)\(\rightarrow\)B:} \{B, AppMK, A, FALSE, MK\}$_{KB}$ \\
\textbf{4'. A\(\rightarrow\)B:} \{B, FALSE, Zero, SKKE\}$_{MK}$ \\
\textbf{5'. B\(\rightarrow\)A:} \{A, TRUE\}$_{MK}$                 \\                           
\textbf{6'. A\(\rightarrow\)B:} \{NA'\}$_{MK}$      \\                
\textbf{7'. B\(\rightarrow\)A:} \{NB'\}$_{MK}$                  \\                       
\textbf{8'.\hspace{0.02cm}A\(\rightarrow\)B:}\hspace{0.02cm}MAC\{3,A,B,NA',NB'\}$_{H(MAC\{A,B,NA',NB'\}_{MK},1)}$        \\
\textbf{9'.\hspace{0.02cm}B\(\rightarrow\)A:}\hspace{0.02cm}MAC\{2,B,A,NB',NA'\}$_{H(MAC\{A,B,NA',NB'\}_{MK},1)}$        \\
\hline
\end{tabular}
\end{table}

\subsection{Proposed Fixed Protocols}
\label{fix}
We propose fixed protocols that use nonces to ensure freshness of the messages and at the same time the keys. 
We make use of the vital principles defined on \cite{Abadi:Needham}. 
The narrations of our proposed solution are given in Table \ref{tab:fix1} and Table \ref{tab:fix2} for Case 1 and Case 2, respectively.

In Case 1, we added the nonce of the initiator (\textbf{NA}) to the first two messages. 
This will ensure that when receiving the second message, A will believe that she is communicating with the TC who knows her nonce and also her private key. 
Note that message 1 can still be replayed but it will be ignored if A does not verify message 2. 
We inserted two more messages before the last message, so that we use nonces of the TC (\textbf{NTC}) and the responder (\textbf{NB}) to avoid replay attacks.
This will ensure that when receiving the fifth message, B will believe that he is communicating with TC who knows his nonce.
Also note that message 3 can still be replayed but the process will be ignored if B does not verify message 5. 

\begin{table}
\caption{Proposed Fix - Case 1}
\label{tab:fix1}
\centering
\begin{tabular}{l}
\hline
\textbf{1. A\(\rightarrow\)TC:} \{TC, AppKey, B, NA\}$_{KA}$   \\
\textbf{2. TC\(\rightarrow\)A:} \{A, AppLK, B, TRUE, NA, LK\}$_{KA}$  \\
\textbf{3. TC\(\rightarrow\)B:} \{B, A, NTC\}$_{KB}$  \\
\textbf{4. B\(\rightarrow\)TC:} \{TC, A, NTC, NB\}$_{KB}$  \\
\textbf{5. TC\(\rightarrow\)B:} \{B, AppLK, A, FALSE, NB, LK\}$_{KB}$  \\
\hline
\end{tabular}
\end{table}
Our solution is also applicable to the leaked MK problem in Case 2. 
Similar to our solution for Case 1, we change the first three messages of Case 2 with five messages that are also given in Table \ref{tab:fix2}.
Not to confuse with the nonces used in SKKE, the nonces we added are called (\textbf{preNA}) and (\textbf{preNB}) in Case 2.

\begin{table}
\caption{Proposed Fix - Case 2}
\label{tab:fix2}
\centering
\begin{tabular}{l}
\hline
\textbf{1. A\(\rightarrow\)TC:} \{TC, AppKey, B, preNA\}$_{KA}$   \\
\textbf{2. TC\(\rightarrow\)A:} \{A, AppMK, B, TRUE, preNA, MK\}$_{KA}$ \\
\textbf{3. TC\(\rightarrow\)B:} \{B, A, NTC\}$_{KB}$ \\
\textbf{4. B\(\rightarrow\)TC:} \{TC, A, NTC, preNB\}$_{KB}$                 \\                           
\textbf{5. TC\(\rightarrow\)B:} \{B, AppMK, A, FALSE, preNB, MK\}$_{KB}$      \\                
\textbf{6. A\(\rightarrow\)B:} \{B, FALSE, Zero, SKKE\}$_{MK}$ \\
\textbf{7. B\(\rightarrow\)A:} \{A, TRUE\}$_{MK}$                 \\                           
\textbf{8. A\(\rightarrow\)B:} \{NA\}$_{MK}$      \\                
\textbf{9. B\(\rightarrow\)A:} \{NB\}$_{MK}$                  \\                       
\textbf{10. A\(\rightarrow\)B:} MAC\{3,A,B,NA,NB\}$_{H(MAC\{A,B,NA,NB\}_{MK},1)}$        \\
\textbf{11. B\(\rightarrow\)A:} MAC\{2,B,A,NB,NA\}$_{H(MAC\{A,B,NA,NB\}_{MK},1)}$        \\
\hline
\end{tabular}
\end{table}
The fix that we propose is a mechanism that suffices to fix the flaws in the original protocol. 
There might be other ways to fix, but this is a solution that simply
works and has proven (by formal verification) to be secure.

Obviously, the proposed solution would come at a particular
cost. Particularly, the
number of messages in each protocol is increased by two, and the usage of
nonces are required. Transmission of more messages means more power
consumption, but for security critical applications (e.g. in Smart
Energy, Commercial Building Automation, etc.)
 this kind of fix which ensures that TC is authenticated to both A and B (i.e. the new LK is not replayed)
is necessary, so the additional messages are inevitable.
Besides the original protocol in Case 2 already has nine messages (whereas the primitive version, Case 1, only has three), which is a proof that in order to
have a sound protocol ZigBee may have longer protocols for the same purpose.
The usage of nonces is not a new cost since it is already in SKKE which is employed by Case 2.
However, the freshness is preserved for only SKKE but not the protocol
itself due to the design mistake of the wrapping protocol.

As we mentioned before, the flaw in End-to-End Application Key Establishment protocol may be visible to an experienced eye but to claim that a fix is flawless, verification using formal methods is crucial.
\emph{Static analysis with \LYSA} is one of the methods that can be used, which has many advantages such as scalability and the guarantee of termination.

\subsection{Formal Verification Details}
\label{zig:formal}
Analysing security protocols without any formal verification method is
not a reliable way to find flaws, nor to guarantee that there are no
flaws. To make our assertions and arguments sound, we use static
analysis to analyse protocols. 
To be finite, this method is computing over-approximations rather than
exact answers, and therefore may lead to false positives. However, when the analysis results tell that the protocol is error-free, then it really is. In other words, no simulation or verification is necessary when the protocols successfully passes static analysis.

The base protocols in Section \ref{zig:keyest} are modelled using \LYSA process calculus and analysed using the \LYSA-tool\footnote{http://www.imm.dtu.dk/English/Research/Language-Based\_Technology/Research/LySa.aspx}. The result supports our claims in Section \ref{flaw}. The base protocols are prone to replay attacks which will cause serious problems in the case of a leaking key.

The proposed protocols in Section \ref{fix} are also modelled analysed in the same way with the base protocol. The result is successful, namely the proposed protocols do not have any flaws.

The settings that we use to implement the \LYSA model and verify in the \LYSA-tool are listed below:
\begin{itemize}
\item we check for the origin and destination addresses in each message (by adding them as prefixes such as in IPv4 or IPv6) 
\item we have the necessary annotations for the encryptions and decryptions
\item we allow legitimate attackers in addition to the illegitimate attackers (by adding appropriate zero indices, namely attacker also shares master key with TC)
\item we model three groups of (infinite) principals so that we can model man-in-the-middle attacks
\item we add an extra message that is encrypted using the session key (to see whether the compromised key can be used)
\item we check all the fields in the messages to have proper values (by pattern matching), except session keys which are newly created (and bound to variables in inputs)
\end{itemize}

To distinguish between old rounds and new rounds of the protocol we apply a new technique in \LYSA. 
We add round indicators to the end of pattern-matched fields in messages and match them in a smart way to distinguish old runs. 
Using this technique, we can investigate replay attacks successfully.

\section{Conclusion}
Analysing protocols is not a trivial issue, and in this work we
presented an analysis method with a detailed application on a new and so called
advanced security protocol that uses secure components. 

In this approach, we have solid benefits in mainly:
\begin{itemize}
\item \emph{solutions always exist and are computed in low polynomial
    time.} This is an important advantage because approaches based on
  model checking cannot always guarantee termination, and besides
  prone to state space explosion problem. Besides the analysis is
  correct with respect to formal operational semantics, which may be hard
  to establish in different approaches such as the ones based on modal
  logic of beliefs (BAN) where the completeness property does not
  generally hold.
\end{itemize}

However, those benefits come with a particular cost:
\begin{itemize}
\item \emph{lack of trace and counter-example.} Due to the nature of
  the analysis, there is no trace and no produced counter-example to
  help flaw discovery. As a result of the over-approximation, false
  positives may occur and manual inspection is required to match the
  reported violations to actual flaws.
\end{itemize}

Another thing we have presented was the usage of protocol analysis in
suggesting a secured version of a flawed protocol. Fixing the flaws
and proposing secure protocols is another non-trivial job. In this manner, we
made use of prudent engineering practices of Gordon and Abadi, and
benefited fruitful discussions with Gavin Lowe. One of the points we
emphasized was the importance of freshness, and the importance of
proper usage of freshness indicators such as nonces, challenges, etc.

We can recapitulate as encryption is not synonymous with security,
and its improper use can lead to errors. The proper use should be 
verified by protocol analysis methods that focus on certain security
properties. Along the way in this study, we discovered and documented general
guidelines about how to use static analysis for protocol
validation. We do believe that such studies are necessary in order
to standardise protocols that live up to their stated
expectations.

\backmatter

\chaptermark{Bibliography}
\renewcommand{\sectionmark}[1]{\markright{#1}}
\sectionmark{Bibliography}


\end{document}